\documentclass[usenatbib,usegraphicx]{mnras}
\usepackage{natbib}
\usepackage{graphicx}	
\usepackage[fleqn]{amsmath}	
\usepackage{amssymb}	
\usepackage{exscale}  
\usepackage{relsize}
\usepackage{xcolor}
\usepackage{longtable}
\PassOptionsToPackage{hyphens}{url}\usepackage{hyperref}  
\hypersetup{
	colorlinks   = true, 
	urlcolor     = blue, 
	linkcolor    = blue, 
	citecolor   = blue 
}

\usepackage{float}
\usepackage{footmisc} 
\usepackage{verbatim}

\newcommand{\crcode}{\texttt{ChN}}
\newcommand{\ka}{K$\alpha$}

\title[Type Ia progenitor for SNR 3C 397]{
Evidence of a Type Ia Progenitor for Supernova Remnant 3C 397}

\author[H. Mart\'{i}nez-Rodr\'{i}guez et al.]{\href{http://orcid.org/0000-0002-1919-228X}{H\'{e}ctor Mart\'{i}nez-Rodr\'{i}guez}$^{1}$\thanks{Email:hector.mr@pitt.edu},
\href{http://orcid.org/0000-0002-1790-3148}{Laura A. Lopez}$^{2,3,4}$,
\href{http://orcid.org/0000-0002-4449-9152}{Katie Auchettl}$^{5,6,4,7}$,\newauthor
\href{http://orcid.org/0000-0003-3494-343X}{Carles Badenes}$^{1}$,
\href{http://orcid.org/0000-0002-7643-0504}{Tyler Holland-Ashford}$^{2}$,
\href{http://orcid.org/0000-0002-7507-8115}{Daniel J. Patnaude}$^{8}$,
\href{http://orcid.org/0000-0002-2899-4241}{Shiu-Hang Lee}$^{9,10}$,\newauthor
\href{http://orcid.org/0000-0003-3462-8886}{Adam R. Foster}$^{8}$,
\href{http://orcid.org/0000-0002-6986-6756}{Patrick O. Slane}$^{8}$\\
$^{1}$Department of Physics and Astronomy and Pittsburgh Particle Physics, Astrophysics and Cosmology Center (PITT PACC), \\ University of Pittsburgh, 3941 O'Hara Street, Pittsburgh, PA 15260, USA\\
$^{2}$Department of Astronomy, The Ohio State University, 140 W. 18th Ave., Columbus, Ohio 43210, USA\\
$^{3}$Center for Cosmology and Astro-Particle Physics, The Ohio State University, 191 West Woodruff Avenue, Columbus, OH 43210, USA\\
$^{4}$DARK, Niels Bohr Institute, University of Copenhagen, Lyngbyvej 2, 2100 Copenhagen, Denmark\\
$^{5}$School of Physics, The University of Melbourne, Parkville, VIC 3010, Australia\\
$^{6}$ARC Centre of Excellence for All Sky Astrophysics in 3 Dimensions (ASTRO 3D)\\
$^{7}$Department of Astronomy and Astrophysics, University of California, Santa Cruz, CA 95064, USA\\
$^{8}$Smithsonian Astrophysical Observatory, 60 Garden St, Cambridge, MA 02138\\
$^{9}$Department of Astronomy, Kyoto University, Kyoto 606-8502, Japan\\
$^{10}$RIKEN, Astrophysical Big Bang Laboratory, 2-1 Hirosawa, Wako, Saitama 351-0198, Japan\\
}

\pubyear{2020}

\begin{document}
\label{firstpage}
\pagerange{\pageref{firstpage}--\pageref{lastpage}}
\maketitle

\begin{abstract}

The explosive origin of the young supernova remnant (SNR) 3C~397 (G41.1$-$0.3) is debated. Its elongated morphology and proximity to a molecular cloud are suggestive of a core-collapse (CC) SN origin, yet recent X-ray studies of heavy metals show chemical yields and line centroid energies consistent with a Type Ia SN. In this paper, we analyze the full X-ray spectrum from $0.7-10$~keV of 3C~397 observed with {\it Suzaku} and compare the line centroid energies, fluxes, and elemental abundances of intermediate-mass and heavy metals (Mg to Ni) to Type Ia and CC hydrodynamical model predictions. Based on the results, we conclude that 3C~397 likely arises from an energetic Type Ia explosion in a high-density ambient medium, and we show that the progenitor was a near Chandrasekhar mass white dwarf. 
\end{abstract}


\section{Introduction} \label{sec:intro}

Supernovae (SNe) are stellar explosions divided into two main categories: core-collapse (CC) and Type Ia SNe. CC SNe arise from massive stars $\gtrsim$8 $M_{\odot}$ that undergo CC when the iron core cannot be supported by nuclear fusion \citep[e.g.,][]{Ib74,ET04,Sm09}. SNe Ia are the thermonuclear explosions of a carbon-oxygen white dwarf (WD) triggered by mass transfer from a companion, which could be a non-degenerate hydrogen- or helium-burning star or another WD (see \citealt{maoz14} for a review). 

SNe are usually typed using their optical spectra around maximum light, days after explosion (see reviews by e.g., \citealt{1997ARA&A..35..309F, 2017hsn..book..195G}), and hundreds of SNe are discovered per year through dedicated surveys (e.g. ASAS-SN: \citealt{2014ApJ...788...48S, Ho17a,Ho17b,Ho17c,Ho18}; ATLAS: \citealt{Ton18}; Pan-STARRs: \citealt{2016arXiv161205243F,2016arXiv161205560C, 2018ApJ...857...51J}, ZTF: \citealt{Smi14,Du18,Kul18} and YSE: \citealt{2019ATel13330....1J}). However, these objects are often too distant ($\sim$1--100~Mpc) to resolve the SN ejecta and the environment of the progenitor star. In this context, supernova remnants (SNRs), which are the leftover structures of SNe that happened hundreds or thousands of years ago, provide a complementary close view of the explosive endpoint of stellar evolution.

In particular, SNRs give valuable information about SN progenitors. At X-ray wavelengths, strong emission lines from shocked SN ejecta can be used to probe the nucleosynthetic products 
\citep[e.g.][]{Ba08a,Park13,Ya15,MR17}. Additionally, their X-ray spectra and morphologies depend on the ejecta, explosion energetics, and the surrounding circumstellar material left by the progenitor
\citep[e.g.][]{Ba03,Ba07,Lop09,Lop11,Pat12,Pat17,Woo17,Woo18}. The most reliable ways to connect SNRs to their progenitors are via detection of an associated pulsar \citep{He68,Ta99} or through light echoes \citep[e.g.,][]{Res05, Res08a,Res08b}. Another approach is to examine the stellar populations surrounding these sources (e.g., \citealt{Ba09,Au18}). 

SNRs can also be typed from their abundance ratios (e.g., \citealt{reynolds07,katsuda18}), morphologies \citep{Lop09,Lop11,Peters13,Lopez18}, Fe-K centroids \citep{Ya14a}, and absorption line studies (e.g., \citealt{hamilton00,fesen17}). However, even with these varied techniques, the explosive origin of some SNRs remain uncertain, such as the Milky Way SNR 3C~397 (G41.1$-$0.3). \citet{Sa00} suggested that 3C~397 arose from a CC SN based on its proximity to molecular clouds and on enhanced abundances of intermediate-mass elements from {\it ASCA} observations. By contrast, \cite{Ya14a} and \cite{Ya15} found that a Type~Ia origin was more likely, given 3C~397's Fe K-shell centroid and its exceptionally high abundances of neutron-rich, stable iron-group elements (Cr, Mn, Ni, Fe; though these elements have been detected in CC SNRs, e.g. \citealt{sato20}). Recent efforts to determine its progenitor metallicity \citep{Da17,MR17} have assumed a Type Ia nature.

Although its exact age is uncertain, 3C 397 is likely ${\sim} \, 1350-5000$ years old \citep[][]{Sa05,Lea16}, implying it is dynamically more evolved than other well-known Type Ia SNRs (e.g. G1.9$+$0.3, 0509$-$67.5, Tycho, Kepler, SN 1006). The X-ray emission from 3C~397 is thermal in nature without a non-thermal component \citep{Ya15}, unlike some other Type Ia SNRs (e.g., G1.9$+$0.3: \citealt{zoglauer15}; Tycho: \citealt{lopez15}). Its X-ray morphology is quite irregular \citep{Sa05}, and \cite{Lee19} suggested that it likely results from interaction with a dense surrounding medium rather than an asymmetric explosion. 

3C~397 is among the class of mixed-morphology SNRs \citep{Rho98}, which are center-filled in X-rays and have a shell-like morphology in the radio. 3C~397 is expanding into a high-density ambient medium (with n$_{\rm amb}\sim 2-5$~cm$^{-3}$; \citealt{Lea16}), with a strong westward gradient. Although this environment is more consistent with a CC SNR, other Type Ia SNRs have signs of interaction: e.g., Tycho may be interacting with a nearby molecular cloud \citep{LeeKT04,Zh16}, N103B has CO clouds along its southeastern edge \citep{sano18}.

In this paper, we aim to better constrain the explosive origin of 3C~397 using the observed emission in the full X-ray band (0.7--10~keV) and comparing it to synthetic SNR spectra generated from both Type Ia and CC explosion models. The paper is organized as follows. In Section~\ref{sec:analysis}, we present the X-ray observations of 3C 397 and the spectral-fitting process. In Section~\ref{sec:discussion}, we compare the observational and synthetic results. Finally, in Section~\ref{sec:conclusions}, we summarize the conclusions.

\section{Observations and data analysis}\label{sec:analysis}

Following the previous studies by \cite{Ya14a} and \cite{Ya15}, we take advantage of the high spectral resolution of the X-ray Imaging Spectrometer (XIS) on board \textit{Suzaku} to measure the centroids and fluxes of all Ly$\alpha$ and \ka \, emission lines. We analyze \textit{Suzaku} observations \texttt{505008010} and \texttt{508001010}, which were taken on 2010 October 24 and 2013 October 30, with total exposure times of 69 and 103 ks, respectively. We use \texttt{HEASOFT} version 2.12 and reprocess the unfiltered public data using the \textit{aepipeline}, the most up-to-date calibration files, and the standard reduction procedures\footnote{\url{http://heasarc.nasa.gov/docs/suzaku/processing/criteria xis.html}}.

We extracted spectra from both the front- (XIS0, XIS3) and back-illumincated (XIS1) CCDs of the entire SNR using an elliptical region and \texttt{XSELECT} version 2.4d. For the background spectrum, we extracted spectra from the full field-of-view of the XIS observations, excluding the calibration regions and the SNR. To generate the redistribution matrix files (RMF) and ancillary response files (ARF), we use the standard \textit{Suzaku} analysis tools \textit{xisrmfgen} and \textit{xisarfgen}, respectively. Due to an error with the ARF analysis pipeline, we are unable to extract a XIS1 spectrum from ObsID \texttt{508001010} and thus do not use this spectrum for our analysis. We also remove (from both the source and background spectra) the contribution of the non-X-ray (i.e., instrumental) background (NXB) that arises from charged particles and hard X-ray \citep{Ta08} interacting with the detectors. To simulate the instrumental background, we use \textit{xisnxbgen} \citep{Ta08} to generate a NXB model and then subtract it from our source and background spectra, similar to what was done in, e.g., \citet{Au15}.

Rather than merging the spectra together, we fit our background-subtracted spectra simultaneously using \texttt{XSPEC}\footnote{\url{https://heasarc.gsfc.nasa.gov/xanadu/xspec/manual/}} \citep[][version 12.10.1]{Ar96} using the standard atomic database. We analyze a broad energy range ($0.7-10.0$~keV) to measure the centroids of all prominent Ly$\alpha$ and \ka\ lines in the 3C~397 data. We model the X-ray continuum using two absorbed bremsstrahlung components (XSPEC model: \textsc{tbabs*(bremss+bremss)}), one soft (with temperature $kT_{\rm s}$) and one hard (with temperature $kT_{\rm h}$). For the emission lines, we include phenomenological Gaussian components with centroid energies, widths, and normalizations tied among the five source spectra. We freeze the column density to $N_{\rm H}=3.5 \times 10^{22}$~cm$^{-2}$ (as derived from our physical model listed in Table \ref{table:3C397_vnei}), similar to \citealt{Ya14a} who used $N_{\rm H} = 3 \times 10^{22}$~cm$^{-2}$ based on the results of \citealt{Sa05}. We adopt solar abundance values from \cite{wilms00}.

\begin{figure}
\includegraphics[width=\columnwidth]{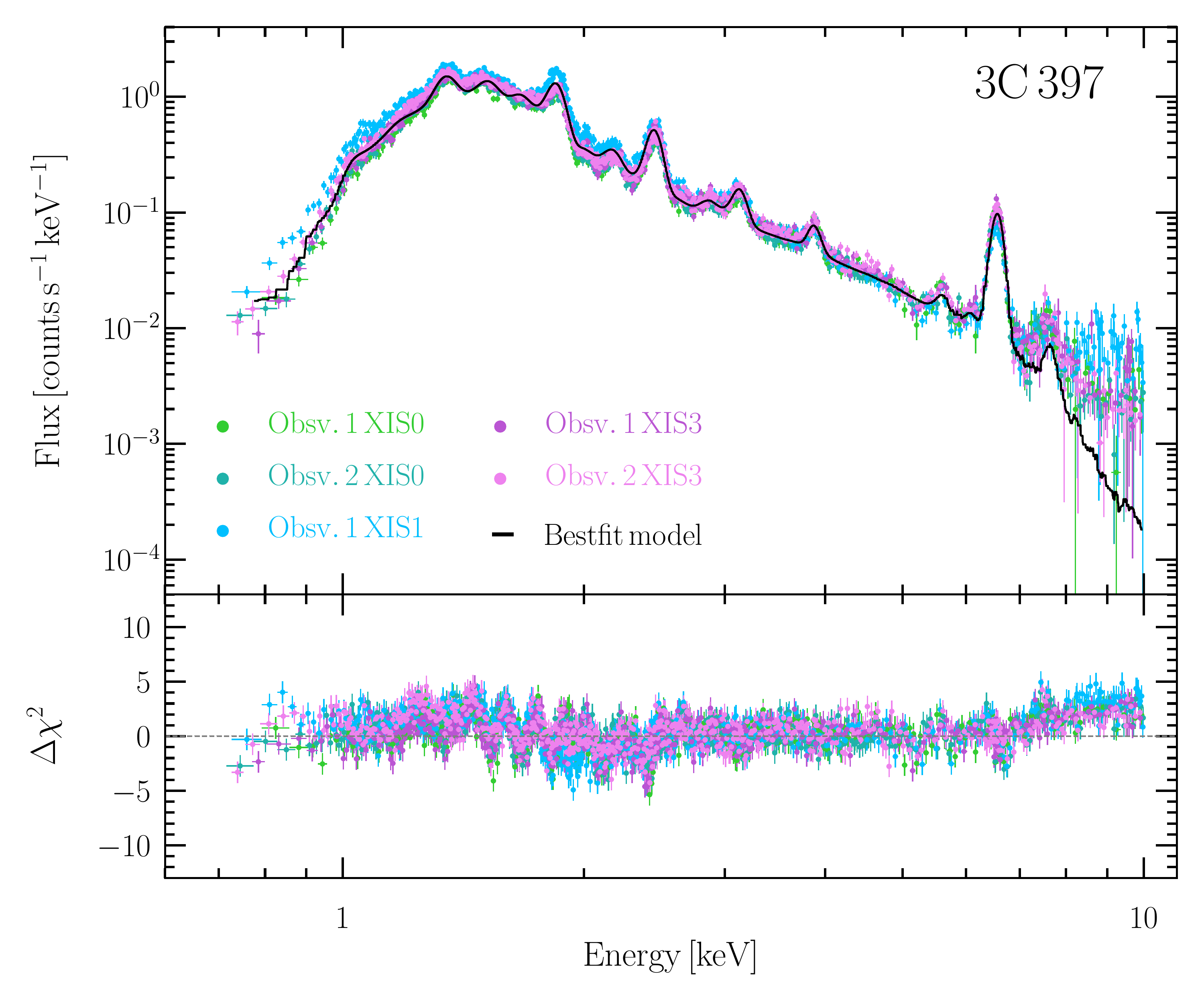}
\caption{The five background-subtracted \textit{Suzaku} X-ray spectra of 3C~397, the best-fit model (as listed in Tables \ref{table:3C397_lines} and \ref{table:3C397_cont}), and the associated residuals.}
\label{fig:Spectra_3C397}
\end{figure}

Tables~\ref{table:3C397_lines} and ~\ref{table:3C397_cont} show the best-fit results, including the centroid energies, line fluxes, and the thermal plasma properties. The spectra and the mean best-fit model for 3C 397 are shown in Figure \ref{fig:Spectra_3C397}. \cite{Ya15}, who modeled the 5--9 keV spectrum from {\it Suzaku} to focus on the Fe-peak elements, derived fluxes for several emission lines common to our analysis: Cr \ka, Mn \ka, Fe \ka,and Ni \ka\ + Fe K$\beta$. Our centroid energies and fluxes are comparable to those measured by \cite{Ya15}, except our Ni \ka\ + Fe K$\beta$ centroid energy is 31~eV lower ($\sim$1.2 $\sigma$) and flux is $\approx$60\% greater than that previous work. This is likely a result of the different continuum temperatures used by \citet{Ya15} and in this analysis ($\sim$2.1 keV versus $\sim$1.6~keV, respectively), and the fact that we used two components rather than one to model the X-ray emission of the SNR.  However, these differences between \citet{Ya15} and our work do not affect our conclusions.

\begin{table}
\begin{center}
\caption{Best-fit model line parameters for the combined \textit{Suzaku} spectra of 3C 397. All uncertainties are 90\% confidence intervals. \label{table:3C397_lines}}
\begin{tabular}{ccc}
\hline
\hline
\noalign{\smallskip}
Transition & Centroid energy & $\rm{\left< Flux \right>}$ \\
\noalign{\smallskip}
& (eV) & $\rm{\left( ph \, cm^{-2} \, s^{-1} \right)}$ \\
\noalign{\smallskip}
\hline
\noalign{\smallskip}
 $\rm{Ne \,\, Ly\alpha}$ & $1027_{-4}^{+5}$ & $\left(1.22_{-1.2}^{+01.3}\right) \times 10^{-2}$ \\
\noalign{\smallskip}
\hline
\noalign{\smallskip}
$\rm{Mg \, K\alpha}$ & $1345_{-1}^{+1}$ & $\left(4.03_{-0.12}^{+0.12}\right) \times 10^{-3}$ \\
\noalign{\smallskip}
\hline
\noalign{\smallskip}
$\rm{Si \, K\alpha}$ & $1853_{-1}^{+1}$ & $\left(1.78_{-0.03}^{+0.03}\right) \times 10^{-3}$ \\
\noalign{\smallskip}
\hline
\noalign{\smallskip}
$\rm{Si \, K\beta}$ & $2218_{-6}^{+4}$ & $\left(9.13_{-0.70}^{+0.70}\right) \times 10^{-5}$ \\
\noalign{\smallskip}
\hline
\noalign{\smallskip}
$\rm{S \, K\alpha}$ & $2454_{-1}^{+1}$ & $\left(5.15_{-0.10}^{+0.10}\right) \times 10^{-4}$ \\
\noalign{\smallskip}
\hline
\noalign{\smallskip}
$\rm{Ar \, K\alpha}$ & $3124_{-4}^{+4}$ & $\left(7.11_{-0.40}^{+0.40}\right) \times 10^{-5}$ \\
\noalign{\smallskip}
\hline
\noalign{\smallskip}
$\rm{Ca \, K\alpha}$ & $3878_{-6}^{+9}$ & $\left(2.01_{-0.20}^{+0.20}\right) \times 10^{-5}$ \\
\noalign{\smallskip}
\hline
\noalign{\smallskip}
$\rm{Cr \, K\alpha}$ & $5601_{-11}^{+12}$ & $\left(1.00_{-0.10}^{+0.10}\right) \times 10^{-5}$ \\
\noalign{\smallskip}
\hline
\noalign{\smallskip}
$\rm{Mn \, K\alpha}$ & $6061_{-13}^{+21}$ & $\left(7.3_{-0.84}^{+0.96}\right) \times 10^{-6}$ \\
\noalign{\smallskip}
\hline
\noalign{\smallskip}
$\rm{Fe \, K\alpha}$ & $6552_{-2}^{+3}$ & $\left(1.39_{-0.03}^{+0.03}\right) \times 10^{-4}$ \\
\noalign{\smallskip}
\hline
\noalign{\smallskip}
$\rm{Ni \, K\alpha \, + \, Fe \, K\beta}$ & $7585_{-12}^{+13}$ & $\left(2.61_{-0.14}^{+0.14}\right) \times 10^{-5}$ \\
\hline
\noalign{\smallskip}
\end{tabular}
\end{center}
\vspace{-5mm}
\end{table}

\begin{table}
\begin{center}
\caption{Best-fit $N_{\rm H}$ and bremsstrahlung components in phenomenological spectral fit. All uncertainties are 90\% confidence intervals. \label{table:3C397_cont}}
\vspace{-5mm}
\begin{tabular}{lcccc}
\hline
\hline
\noalign{\smallskip}
$N_{\rm{H}}$ & $kT_{\rm{s}}$ & Norm$_{\rm{s}}^{\rm a}$ & $kT_{\rm{h}}$ & Norm$_{\rm{h}}^{\rm a}$ \\
\noalign{\smallskip}
(10$^{22}$ cm$^{-2}$) & (keV) &  & (keV) & \\
\noalign{\smallskip}
\hline
\noalign{\smallskip}
3.49 [frozen] & 0.24$\pm$0.01 & $14.0_{-0.6}^{+0.7}$ & 1.60$\pm$0.03 & 0.02$\pm$0.001 \\
\noalign{\smallskip}
\hline
\end{tabular}
\end{center}
$^{\rm a}$ The normalizations Norm$_{\rm s}$ and Norm$_{\rm h}$ are given in units of ($10^{-14}/4 \pi D^{2}$) $\int n_{\rm e} n_{\rm H} \, dV$ (cm$^{-5}$), where $D$ is the distance to the source (cm), $n_{\rm e}$ and $n_{\rm H}$ are the electron and hydrogen densities ($\mathrm{cm}^{-3}$), respectively. 
\end{table}

\section{Results and Discussion}\label{sec:discussion}

\subsection{Explosive Origin Constraints from Line Ratios and Centroid Energies}

\begin{figure*}
\centering
\includegraphics[width=\textwidth]{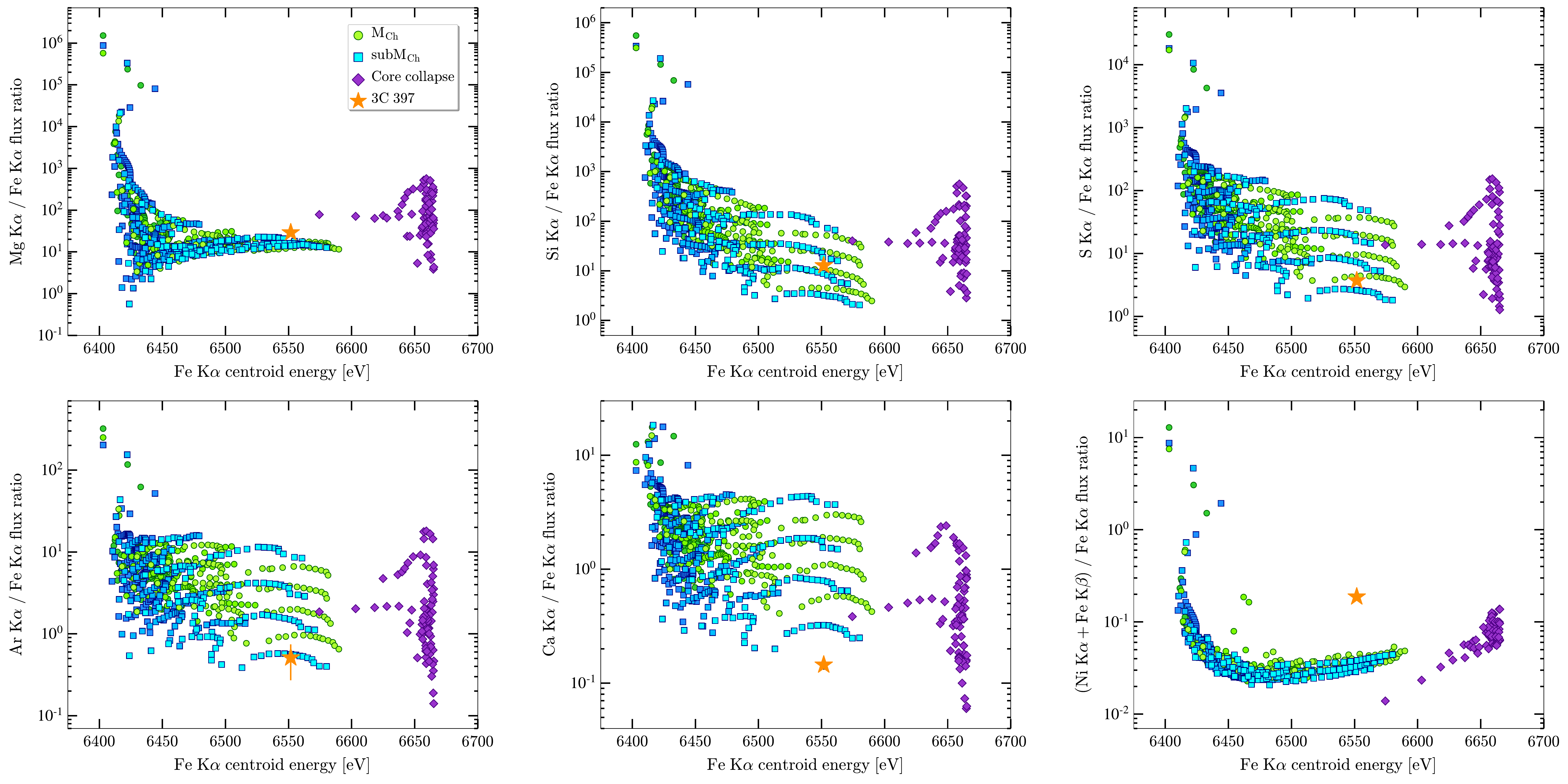}
\vspace{-5mm}
\caption{Emission line fluxes normalized to the Fe \ka\, flux versus Fe \ka\, centroid energy for the $M_{\rm Ch}$ models (green circles), sub-$M_{\rm Ch}$ models (blue squares), CC models (purple diamonds; from \citealt{Pat15}), and observed from 3C~397 (orange star). Lighter shades of blue and green correspond to greater ambient densities.}
\label{fig:Flux_Ratios_3C397}
\end{figure*}

We compare the observational fluxes reported in Table \ref{table:3C397_lines} with theoretical models for the X-ray spectra of both Type Ia and CC SNRs. In contrast to \cite{Ya15}, who focused on the $M_{\rm{Mn}} / M_{\rm{Fe}}$ and $M_{\rm{Ni}} / M_{\rm{Fe}}$ mass ratios, we examine a broader range of metals that includes both intermediate-mass and Fe-peak elements. 

We use a grid of synthetic X-ray spectra \citep[see][for a detailed explanation]{MR18} for Type Ia SN models that assume a progenitor with a metallicity of $Z = 0.009$ ($\approx0.64~Z_{\sun}$) which is expanding into the SNR phase, similar to that used in previous studies of Type Ia SNRs \citep[e.g.,][]{Ba03,Ba05,Ba08b,MR18}. We analyze the synthetic spectra from the X-ray emitting ejecta in $M_{\rm Ch}$ and sub-$M_{\rm Ch}$ models \citep{Br19}, adopting three uniform ambient medium densities: $\rho_{\rm amb}$ = $10^{-24}$, $2\times10^{-24}$, and $5\times10^{-24}$~g cm$^{-3}$  (corresponding to ambient number densities of $n_{\rm amb} = 1$, 2, and 5.0~cm$^{-3}$). These values are consistent with the estimated densities around 3C~397 of $n_{\rm amb} \ {\sim} \ 2-5 \, \rm{cm}^{-3}$ \citep{Lea16}.

In addition, we consider synthetic X-ray spectra from single-star CC explosion models (specifically, models s25D and s12D from \citealt{heger10}) and adopted in previous SNR studies \citep{Lee14,Pat15}. In total, we produce eight SNR models using two sets of mass-loss rates and wind velocities ($10^{-5} \, M_{\odot}\rm{ \, yr^{-1}}$, 10 $\rm{km \, s^{-1}}$ and $2 \times 10^{-5} \, M_{\odot}\rm{ \, yr^{-1}}$, 20 $\rm{km \, s^{-1}}$) and four CC SN ejecta profiles. The four ejecta profiles are from two stars with initial masses of 12 and 25 $M_{\odot}$  (that lose $\sim$3 and 13 $M_{\odot}$, respectively, by the onset of CC), a 6 $M_{\odot}$ He star enclosed in a 10~$M_{\odot}$ H envelope (tailored to mimic SN~1987A), and a 18 $M_{\odot}$ main-sequence star with a mass-loss of 15 $M_{\odot}$ by CC (matched to Type IIb SN~1993J). We note that while this set of CC SNR models is not comprehensive, it is diverse enough to be representative and has been shown to provide a good match to the bulk dynamics of most CC SNRs \citep{Pat15,Pat17}.

We calculate centroid energies and line fluxes from the unconvolved, differential photon fluxes of these SNR model spectra using Equations 2 and~3 of \citet{MR18}. For each transition, we select the energy integration range from the 3$\sigma$ limit of the corresponding Gaussian profile in the convolved {\it Suzaku} spectra.

Figure \ref{fig:Flux_Ratios_3C397} shows the emission line ratios (relative to Fe \ka) versus the Fe \ka\ centroid energy derived from the synthetic spectra. We normalize the ratios relative to Fe \ka\ because that line is detected in many SNRs and is useful to characterize SN progenitors \citep{Ya14a,Pat15,PatB17,MR18}. In SNRs, the Fe \ka\, flux is sensitive to the electron temperature and ionization timescale, and the Fe \ka\, centroid energy is an excellent tracer of the mean charge state of Fe \citep{Vi12,Ya14b,Ya14a}. As a consequence, the latter can be used to distinguish whether SNRs arise from Type Ia and CC SNe \citep{Ya14a}, with the former having centroids $<$6550~eV and the latter having centroids $>$6550~eV. \footnotetext{We note that one possible exception is the SNR W49B, which has a Fe \ka\ centroid of 6663$\pm$1~eV reported by \cite{Ya14a}, and it is debated whether the originating explosion was a Type Ia \citep{zhou18} or CC SN \citep{lopez09a,lopez13}.} We find that 3C~397 has a  Fe \ka\ centroid of 6552$^{+3}_{-2}$~eV, consistent with the value derived by \cite{Ya14a} of 6556$^{+4}_{-3}$~eV and at the boundary that distinguishes Type Ia from CC progenitors.

We find that at the measured value of the Fe \ka\ centroid, the observed line flux rations derived for 3C~397 are broadly consistent with the Type Ia $M_{\rm Ch}$ and sub-$M_{\rm Ch}$ models and are incompatible with the CC models. The Mg/Fe flux ratio of 3C~397 is $\sim$50\% greater than our Type Ia model predictions, but the Si/Fe, S/Fe, and Ar/Fe flux ratios are consistent with both Type Ia scenarios as long as $\rho_{\rm amb} \, \gtrsim \, (2.0-5.0) \times 10^{-24} \, \rm{g \, cm^{-3}}$.  

The Ca/Fe flux ratio in 3C~397 is $\sim$2.5$\times$ below our model predictions. Previous studies comparing the derived emission properties of Ca \ka\ to hydrodynamical models have found similar inconsistencies. For example, \cite{MR17} showed that the Ca/S mass ratio measured from X-ray spectra of Type Ia SNRs cannot be reproduced with the standard reaction rates used in most SN Ia explosion models. Both \citet{Ya15} and \cite{MR17} pointed out that the (Ni K$\alpha$+Fe K$\beta$)/FeK$\alpha$ flux ratio is exceptionally large for a Type Ia SNR. \citet{MR17} showed that this large ratio is suggestive of a high-metallicity progenitor, which may also explain the anomalous Ca/Fe ratios seen for 3C~397. 

We note that discrepancies between the observed values and the models may be due to well-documented challenges in comparing simple explosion models to an entire X-ray spectrum. For example, one-dimensional hydrodynamic models cannot account for variations in interstellar absorption, non-thermal contribution, and background across the SNR (see \citealt{Ba03} and \citealt{Ba06}).

\begin{figure*}
\centering
\includegraphics[width=\textwidth]{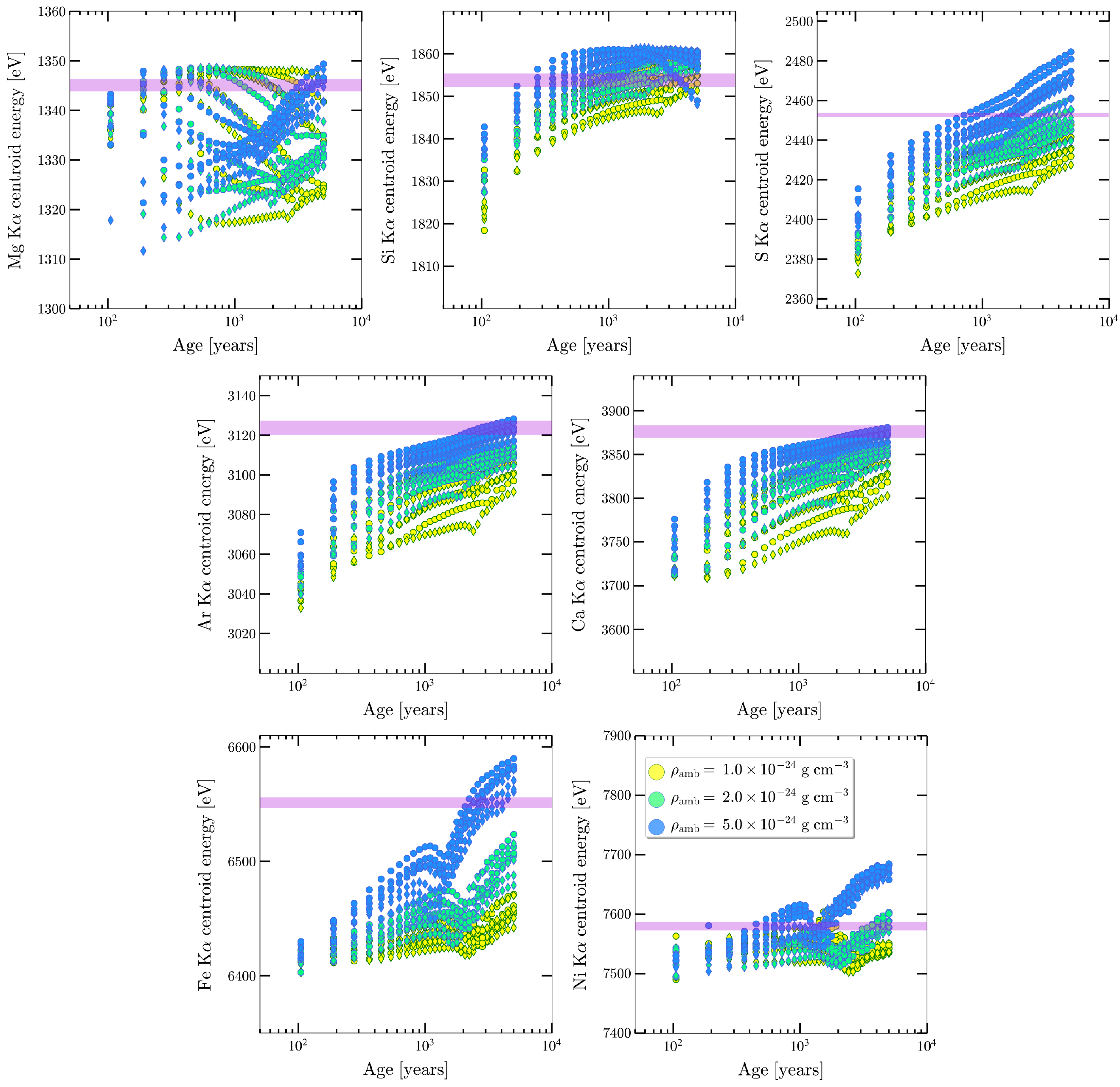}
\vspace{-5mm}
\caption{Emission-line centroid energies versus expansion age for the $M_{\rm Ch}$ models (circles), sub-$M_{\rm Ch}$ models (diamonds) of different ambient densities: $\rho_{\rm amb}$=[1.0, 2.0, 5.0] $\times10^{-24}$ g cm$^{-3}$. The shaded purple region corresponds to the best-fit emission-line centroid  energy of 3C397 as derived in Table \ref{table:3C397_lines}.} 
\label{fig:Centroids_3C397}
\end{figure*}

Figure \ref{fig:Centroids_3C397} shows the theoretical and observational centroid energies for the transitions depicted in Figure \ref{fig:Flux_Ratios_3C397}. These centroids tend to have higher energies for greater expansion ages and ambient densities. For Mg and Si, the observed values in 3C~397 are consistent with both $M_{\rm Ch}$ and sub-$M_{\rm Ch}$ Type Ia models of medium ambient densities and a wide range of ages ($\gtrsim 200-5000$ years). The centroid energies of S, Ar, Ca, Fe and Ni are more consistent with the highest ambient medium densities ($\rho_{\rm amb} = 5.0\times10^{-24}$~g~cm$^{-3}$), suggesting that 3C~397 is in a dense environment, consistent with that found by \cite{Lea16} and its irregular morphology \citep{Lee19}. While the centroid energies of S, Ar, and Ni can occur over a wide range of ages ($\gtrsim 700-5000$ years), the Ca and Fe centroids set the most stringent constraints. Our results suggest that 3C~397 has an age between 2000--4000 years, consistent with (but more constraining than) estimates reported in the literature (1350--5300 years: \citealt{Sa00,Sa05,Lea16,2018ApJ...866....9L}).

\begin{figure*}
\centering
\includegraphics[width=\textwidth]{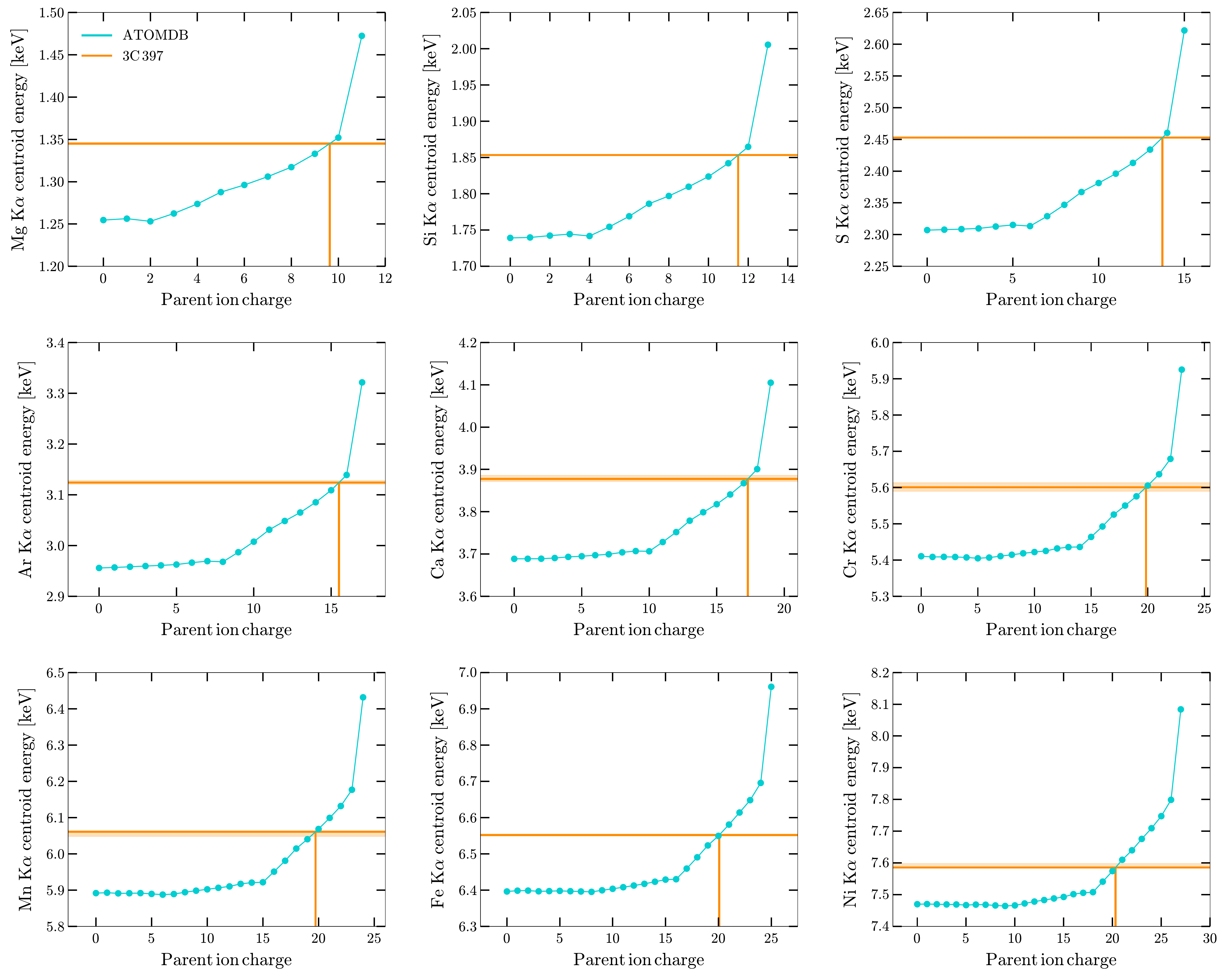}
\vspace{-3mm}
\caption{\ka\, emission line centroid energies versus ion parent charge (charge number) from the \textit{AtomDB} database. The best-fit values for 3C 397 (Table~\ref{table:3C397_lines}) are shown as orange shaded regions. The Fe-peak elements are more ionized than the intermediate-mass elements and are more consistent with Type Ia progenitor.}
\label{fig:Centroids_ATOMDB}
\end{figure*}

To further explore the ionization state of the plasma, we extract the centroid energy as a function of the parent ion charge for all of the observed \ka\ transitions listed in Table~\ref{table:3C397_lines} (Mg, Si, S, Ar, Ca, Cr, Mn, Fe and Ni) using the \textit{AtomDB} database \citep{Fo12, Fo14}. Figure \ref{fig:Centroids_ATOMDB} shows these centroid energies and the values measured for 3C 397, including the corresponding ionization state for each transition. The derived centroids suggest that the plasma of 3C~397 is highly (but not fully) ionized, and the charge number of the Fe-peak elements saturates at an ion charge of 20. These values are at the extreme end of observations of other Type Ia SNRs, though they are still lower than those found for CC SNRs \citep[c.f., Figure 1 of][]{Ya14b}, supporting the Type Ia progenitor origin of 3C~397.

\subsection{Explosive Origin Constraints from Plasma Ionization State and Metal Abundances}

To further probe the explosive origin of 3C~397, we search for evidence of overionization (recombination) by fitting the $0.7-10$~keV spectrum with multiple non-equilibrium ionization (NEI) model components ({\sc vvrnei}). Overionization is a signature of rapid cooling that causes the ions to be stripped of more electrons than expected for the observed electron temperature of the plasma. This rapid cooling could arise from thermal conduction \citep{2002ApJ...572..897K}, adiabatic expansion \citep{1989MNRAS.236..885I}, or interaction with dense material \citep{2005ApJ...630..892D}. To date, overionization has only been detected in mixed-morphology SNRs (e.g., W49B: \citealt{Ka05,Oz09,Mic10,Lo13}; IC443: \citealt{Ka02,Ka05,2009ApJ...705L...6Y,Oh14, Ma17}), many of which have been classified as CC SNRs based on their elemental abundances, their morphologies, and the dense material in their environments.

We find that the ejecta emission of 3C 397 is best described by an underionized plasma, where the temperature of the electrons is greater than the ionization temperature, contrary to an overionized plasma. However, we note that the absence of overionization does not exclude a CC origin, since many CC SNRs (such as Cassiopeia~A: \citealt{Hug00}) are also underionized. 

Finally, we aim to constrain the explosive origin of 3C~397 based on the abundance ratios from the $0.7-10$~keV {\it Suzaku} spectra. \cite{Sa05} analyzed a 66~ks {\it Chandra} observation of 3C~397 and found that the emission was ejecta-dominated and best fit by two NEI plasma components. However, due to low signal-to-noise, the derived metal abundances (e.g., of Si and Fe) were not well constrained. Subsequent work using the {\it Suzaku} observations of 3C~397 analyzed specific energy bands (e.g., 2--5 keV: \citealt{MR17}; 5--9 keV: \citealt{Ya14a,Ya15}) rather than the full X-ray spectrum. 

We find that three absorbed NEI plasma (\textsc{tbabs*(nei+vnei+vvnei)}) components best describe the spectra of 3C~397 (see Table~\ref{table:3C397_vnei} for the best-fit parameters). Here, we let the column density $N_{\rm H}$, ionisation timescale $\tau$, normalisation, and temperatures of each NEI component be free parameters. Due to the strong emission lines from Si, S, Ar, Ca, Cr, Mn, Fe, and Ni, the abundances of these elements were also allowed to vary, while all other elements in each component were set to solar. We find that the two hottest components have super-solar abundances and are associated with ejecta, whereas the coolest component has ISM (solar) abundances. The ionisation timescale of the ISM component was frozen to $\tau = 5\times10^{13}$~s~cm$^{-3}$ as this parameter was unconstrained. We also add three Gaussians, two with centroid energies of 1.01$\pm$0.03 and 1.23$\pm$0.02~keV to compensate for large residuals that correspond to Ne Ly$\alpha$ and Ne {\sc x} (or Fe {\sc XXI}),  respectively \citep{Fo12, Fo14}. It is possible that the Ne could be at a different temperature or ionization to the rest of the plasma, causing it to not be fully captured by our NEI models.  The third Gaussian, with a centroid energy of 6.43$\pm$0.01~keV, accounts for the low centroid of the Fe \ka\ emission line in 3C~397, which is less than the $\sim$6.7 keV Fe peak energy assumed in the vnei/vvnei components. With the addition of the three Gaussians, our best fit yields $\chi^{2}_{\rm reduced} = 1.89$.

\begin{table}
\caption{The best-fit parameters from physical model of spectra. All uncertainties are 90\% confidence intervals.}
\label{table:3C397_vnei}
\begin{center}
\begin{tabular}{lcc}
\hline
\hline
Component & Parameter & Value \\
\hline
tbabs         & N$_{\rm H}$ ($\times10^{22}$ cm$^{-2}$) & 3.49$_{-0.04}^{+0.02}$ \\
  \hline
nei         & $kT_{\rm s}$ (keV) & 0.22$\, \pm \,0.02$\\
              & $\tau$ ($\times 10^{13}$ s cm$^{-3}$) &5.00 [frozen] \\
              & normalization$^{\rm a}$ ($\times10^{-1}$) & 7.65$\pm$0.6\\
  \hline
vnei          & $kT_{\rm s}$ (keV) & 0.58$\pm$0.01\\
              & Si & 3.37$_{-0.09}^{+0.11}$ \\
              & S & 4.28$_{-0.14}^{+0.10}$  \\
              & Ar & 6.56 $_{-1.00}^{+0.60}$\\
              & Ca & 12.4$_{-3.0}^{+2.0}$\\
              & $\tau$ ($\times 10^{11}$ s cm$^{-3}$)& 4.07$_{-0.49}^{+0.59}$  \\
              & normalization$^{\rm a}$ ($\times10^{-2}$)& 5.8$\pm$0.1 \\
  \hline
vvnei       & $kT_{\rm h}$ (keV) & 1.89$\, \pm \,0.03$ \\
              & Cr & 25.3$_{-4.2}^{+4.1}$ \\
              & Mn & 57.7$_{-12}^{+9.3}$ \\
              & Fe & 13.2$_{-0.7}^{+0.4}$\\
              & Ni & 62.7$_{-6.2}^{+3.3}$ \\
              & $\tau$ ($\times 10^{11}$ s cm$^{-3}$)& 1.05$_{-0.1}^{+0.4}$  \\
              & normalization$^{\rm a}$ ($\times10^{-2}$)& 1.7$\pm$0.1\\
\hline
            & $\chi^{2}_{\rm reduced}$ & 1.89 \\
\hline
\end{tabular}
\end{center}
$^{\rm a}$ The normalizations are given in units of ($10^{-14}/4 \pi D^{2}$) $\int n_{\rm e} n_{\rm H} \, dV$ (cm$^{-5}$), where $D$ is the distance to the source (cm), $n_{\rm e}$ and $n_{\rm H}$ are the electron and hydrogen densities ($\mathrm{cm}^{-3}$), respectively. 
\end{table}

We find super-solar abundances of metals in the two hottest components, with some elements (e.g., Cr, Mn, Ni) extremely enhanced relative to others (e.g., Fe). Super-solar abundances and ejecta-dominated emission is common among mixed-morphology SNRs \citep[e.g.,][]{La06,Uc12,Au14,Au15}. However, the extreme abundances of the Fe peak elements \citep{Ya15,MR17} suggested that 3C~397 arose from a Chandrasekhar mass progenitor that produced significant neutron-rich material during the explosion. 

We calculate the X-ray emitting mass swept-up $M_{\rm X}$ by the forward shock of 3C~397 using \hbox{$M_{X} = 1.4 m_{\rm H} n_{\rm H} f V$}, where $m_{\rm H}$ and $n_{\rm H}$ is the mass and number density of hydrogen, $V$ is the volume, and $f$ is the filling factor. We adopt a distance to the SNR of $D = 8.5$~kpc \citep{rana18} and a radius of 2.5\arcmin\ $\approx$ 6.2~pc. Based on the best-fit normalization of the ISM plasma and assuming $n_{\rm e} = 1.2 n_{\rm H}$, we find $n_{\rm H} = 4.4$~cm$^{-2}$, consistent with previous measurements in the literature \citep{Lea16} and the results from Section~\ref{sec:discussion}. The corresponding $M_{\rm X}$ is 148 $d^{5/2}~f^{1/2}$~$M_{\sun}$ (where $d$ is the distance scaled to 8.5~kpc), suggesting the SNR is in the Sedov-Taylor phase.  

Assuming that the reverse shock has heated all of the ejecta, we estimate the mass of ejecta by summing the mass of each element $M_{\rm i}$ given the measured abundances: $M_{\rm i}=[(a_{\rm i}-1)/1.4](n_{\rm i}/n_{\rm H})(m_{\rm i}/m_{\rm H})M_{\rm tot}$. Here $a_{\rm i}$ is the abundance of element $i$ listed in Table \ref{table:3C397_vnei}, $m_{\rm i}$ is the atomic mass of element $i$, $n_{\rm i}/n_{\rm H}$ is its ISM abundance relative to hydrogen, and $M_{\rm tot}$ is the total mass of the ejecta thermal components. Based on the abundances in Table \ref{table:3C397_vnei}, we find that an ejecta mass of $\sim1.22\,d^{5/2}f^{1/2} M_{\sun}$, consistent with a Type Ia explosive origin.

From the abundances listed in in Table~\ref{table:3C397_vnei}, we calculate the mass ratios $M_{\rm Fe}/M_{\rm S} = 11.7^{+0.3}_{-0.2}$ and $M_{\rm Si}/M_{\rm S} = 1.05\pm0.01$. Here we have assumed that all ejecta have been shocked. These values are consistent with those of our most energetic sub-$M_{\rm Ch}$ Type Ia SN models, whereas the $M_{\rm Fe}/M_{\rm S}$ from our fits is $\gtrsim5\times$ the predictions from CC models of \cite{Pat15} and \cite{Suk16a}. The $M_{\rm Si}/M_{\rm Fe}$ ratio can be used to constrain the white dwarf progenitor mass \citep{2018ApJ...857...97M}. We find $M_{\rm Si}/M_{\rm Fe} = 0.09\pm0.002$ for 3C~397, which corresponds to a $\sim$1.06-1.15~$M_{\sun}$ white dwarf from the Bravo models presented in \cite{2018ApJ...857...97M}. \cite{Ya15} ruled out sub-$M_{\rm Ch}$ models for 3C~397 based on the Ni/Fe and Mn/Fe mass ratios, and our result is consistent with that conclusion.

\section{Conclusions}\label{sec:conclusions}

We analyze the {\it Suzaku} X-ray observations of SNR 3C 397 to constrain its explosive origin. We measure the centroid energies and line fluxes using a phenomenological model, and we compare the values to those derived from synthetic spectra produced by Type Ia and CC explosion models. We find 3C~397 is most consistent with a Type Ia SN scenario that occurred in a high-density ambient medium ($\rho_{\rm amb} \gtrsim$2--5$\times10^{-24}$~g~cm$^{-3}$) $\approx$2000--4000 years ago. We model the $0.7-10$~keV X-ray spectra using multiple NEI components, and we find that the ejecta are underionized and have super-solar abundances consistent with a Type Ia origin. Finally, we calculate elemental mass ratios and compare to Type Ia and CC models. We show that these ratios are consistent with the former, and the $M_{\rm Si}/M_{\rm Fe}$ ratio suggests a white dwarf progenitor near $M_{\rm Ch}$. \\

\section*{Acknowledgments}

Support for this work has been provided by the Chandra Theory award TM8-19004X. H.M.-R. acknowledges funding as a CCAPP Price Visiting Scholar, supported by the Dr. Pliny A. and Margaret H. Price Endowment Fund. H.M.-R. also acknowledges support from the NASA ADAP grant NNX15AM03G S01,  a PITT PACC, a Zaccheus Daniel and a Kenneth P. Dietrich School of Arts
\& Sciences Predoctoral Fellowship. KAA is supported by the Danish National Research Foundation (DNRF132). This research has made use of NASA's Astrophysics Data System (ADS, \url{http://adswww.harvard.edu/}).
Parts of this research were supported by the Australian Research Council Centre of Excellence for All Sky Astrophysics in 3 Dimensions (ASTRO 3D), through project number CE170100013.

{\it Software}: \texttt{FTOOLS} \citep{Bl95}, \texttt{XSPEC} \citep{Ar96}, \texttt{SAOIMAGE DS9} \citep{Jo03}, \crcode\ \citep{Ell07,Pat09,Ell10,Pat10,Cas12,Lee14,Lee15}, \texttt{Matplotlib} \citep{Hun07}, \texttt{IPython} \citep{PeG07}, \texttt{Numpy} \citep{vaW11}, \texttt{PyAtomDB} (\url{http://atomdb.readthedocs.io/en/master/}), \texttt{Astropy} \citep{Astro13,Astro18}, \texttt{Python} (\url{https://www.python.org/}), \texttt{SciPy} (\url{https://www.scipy.org/}).

\bibliographystyle{mnras_bib}
\bibliography{Paper_3C397}

\end{document}